\documentclass[sigconf]{acmart}

\copyrightyear{2026}
\acmYear{2026}
\setcopyright{cc}
\setcctype{by}
\acmConference[ICS '26]{2026 International Conference on Supercomputing}{July 06--09, 2026}{Belfast, United Kingdom}
\acmBooktitle{2026 International Conference on Supercomputing (ICS '26), July 06--09, 2026, Belfast, United Kingdom}
\acmDOI{10.1145/3797905.3807844}
\acmISBN{979-8-4007-2522-7/2026/07}
\usepackage{enumitem}
\usepackage{xfrac}
\usepackage{xspace}
\usepackage{subfig}

\newcommand{\NAME}{GPIR\xspace}

\newcommand{\mathDB}{\mathtt{DB}}
\newcommand{\DB}{$\mathDB$\xspace}
\newcommand{\ciphertext}{$\mathtt{ct}$\xspace}

\newcommand{\evk}{$\mathtt{evk}$\xspace}
\newcommand{\ciphertexts}{$\mathtt{ct}$s\xspace}

\newcommand{\evks}{$\mathtt{evk}$s\xspace}
\newcommand{\mathsubs}{\mathrm{Subs}}
\newcommand{\subs}{$\mathsubs$\xspace}

\begin{document}

\title{GPIR: Enabling Practical Private Information Retrieval with GPUs}


\author{Hyesung Ji}
\orcid{0009-0009-9288-159X}
\affiliation{%
 \institution{Seoul National University}
 \city{Seoul}
 \country{South Korea}
}
\email{kevin5188@snu.ac.kr}

\author{Hyunah Yu}
\orcid{0009-0008-4949-5688}
\affiliation{%
 \institution{Seoul National University}
 \city{Seoul}
 \country{South Korea}
}
\email{yhyuna@snu.ac.kr}

\author{Jongmin Kim}
\orcid{0000-0003-2937-3073}
\affiliation{%
 \institution{Seoul National University}
 \city{Seoul}
 \country{South Korea}
}
\email{jongmin.kim@snu.ac.kr}

\author{Wonseok Choi}
\orcid{0009-0004-0941-4805}
\affiliation{%
 \institution{Seoul National University}
 \city{Seoul}
 \country{South Korea}
}
\email{wonseok.choi@snu.ac.kr}

\author{G. Edward Suh}
\orcid{0000-0001-6409-9888}
\affiliation{%
 \institution{NVIDIA}
 \city{Westford}
 \state{MA}
 \country{USA}
}
\affiliation{%
 \institution{Cornell University}
 \city{Ithaca}
 \state{NY}
 \country{USA}
}
\email{esuh@nvidia.com}

\author{Jung Ho Ahn}
\orcid{0000-0003-1733-1394}
\affiliation{%
 \institution{Seoul National University}
 \city{Seoul}
 \country{South Korea}
}
\email{gajh@snu.ac.kr}

\renewcommand{\shortauthors}{Hyesung Ji et al.}

\begin{abstract}

Private information retrieval (PIR) allows private database queries; however, it is hindered by intense server-side computation and memory traffic.
Numerous modern lattice-based PIR protocols consist of three phases: ExpandQuery (expanding a query into encrypted indices), RowSel (encrypted row selection), and ColTor (recursive ``column tournament'' for final selection).
ExpandQuery and ColTor primarily perform number-theoretic transforms (NTTs), whereas RowSel reduces to large-scale independent matrix--matrix multiplications (GEMMs).
GPUs are well suited for these tasks when combined with multi-client batching, which is necessary for high throughput.
However, batching fundamentally reshapes the performance bottlenecks: while it amortizes database access costs, it expands working sets beyond the L2 cache capacity, causing divergent memory access behavior and excessive DRAM traffic.

We present GPIR, a GPU-accelerated PIR system that rethinks kernel design, data layout, and execution scheduling. 
We introduce a stage-aware hybrid execution model that dynamically switches between operation-level kernels, which execute each primitive operation separately, and stage-level kernels, which fuse all operations within a stage into a single kernel to maximize on-chip data reuse.
For RowSel, we resolve the mismatch between NTT-driven layouts and tiled GEMMs using a transposed-layout design with fine-grained pipelining.
We further extend GPIR to multi-GPU systems, scaling throughput and database capacity with negligible communication overhead. 
GPIR achieves up to 297.2$\times$ higher throughput than PIRonGPU, the state-of-the-art GPU implementation.
\end{abstract}


\maketitle

\section{Introduction}
\label{sec:intro}

Private information retrieval (PIR) enables a client to retrieve a record from a public database (\DB) hosted on a server without revealing to anyone which record is accessed.
PIR enables fundamentally new classes of privacy-preserving \DB services, exemplified by Apple's private visual search~\cite{AppleHE}, private DNS resolution~\cite{sigcomm-2023-pdns}, and private reads in blockchain systems~\cite{ethereum-pir}.

Despite these promising applications, deploying PIR at scale remains challenging due to its substantial server-side computation and memory access demands.
Modern PIR protocols~\cite{usenixsec-2023-simplepir, usenixsec-2024-ypir, ccs-2021-onionpir, iacr-2025-onionpirv2, crypto-2024-hintless, sp-2018-sealpir, popets-2016-xpir, sp-2022-spiral} rely on lattice-based homomorphic encryption (HE)~\cite{stoc-2009-gentry} to provide strong privacy guarantees.
HE is a core enabler of PIR, allowing servers to perform encrypted record selection without learning the queried \DB index.
However, this capability comes at a high cost: each query requires computation over the entire \DB~\cite{jacm-1998-chorpir} and incurs substantial overhead from homomorphic ciphertext operations~\cite{access-2021-demystify}.
For GB-scale \DB sizes, these overheads quickly become intractable, leading to multi-second query latencies~\cite{usenixsec-2024-ypir} and severely limiting the practicality of the aforementioned services.

GPUs~\cite{ieee-micro-2021-gpu, ASPLOS-2024-gpupir, usenixsec-2022-gpupir} offer a compelling opportunity for accelerating PIR, as their massive parallelism and high memory bandwidth are well suited for accelerating core PIR operations, such as number-theoretic transforms (NTTs) and point-wise parallel arithmetic.

Modern lattice-based PIR protocols execute in three distinct phases: (1) ExpandQuery, which expands a client query into a vector of encrypted indices; (2) RowSel, which performs homomorphic row selection from the \DB; and (3) ColTor, a recursive ``column tournament'' to select the final record. To achieve high throughput, servers employ multi-client batching, where queries from numerous clients are processed concurrently to amortize \DB access costs.

We observe that multi-client batching reshapes PIR performance on GPUs, creating a classic systems trade-off.
Batching increases arithmetic intensity and enables RowSel---a phase that reduces to independent matrix--matrix multiplications (GEMMs) and is otherwise severely constrained by memory bandwidth---to approach the GPU roofline's balanced ridge point~\cite{ispass-2026-theodosian}.
However, batching dramatically expands the working sets of ExpandQuery and ColTor, pushing memory demands beyond the GPU's L2 cache capacity.
When these working sets exceed the hardware threshold, performance collapses due to excessive DRAM traffic.
Na\"ively applying existing GPU kernels fails to resolve these bottlenecks, which we term the \emph{cache-capacity wall}.

This paper presents \NAME, \emph{a GPU-accelerated PIR system optimized for scalable, batched execution}.
Based on a detailed roofline analysis, we introduce a stage-aware hybrid execution model that dynamically selects kernel granularity for each stage: operation-level kernels maximize occupancy when the working set is L2-resident, while stage-level kernels fuse operations within a stage to minimize DRAM traffic once the cache capacity is exceeded.
Further, we identify a performance gap in RowSel caused by a structural mismatch between NTT-driven HE data layouts and GEMM-friendly layouts required for high-throughput tiling.
We resolve this with a transposed-layout RowSel design and fine-grained pipelining to overlap data transpositions with GEMMs.
Finally, we extend \NAME to multi-GPU systems, leveraging \DB sharding and high-bandwidth interconnects to scale throughput and support \DB exceeding single-GPU memory limits with negligible communication overhead.

The main contributions of the paper are as follows:
\begin{itemize}[leftmargin=*, nosep]
    \item \textbf{Architectural analysis of batched PIR:} 
    We characterize the ``cache-capacity wall'' in batched PIR and show how batching shifts bottlenecks to DRAM traffic, motivating a stage-aware hybrid execution model for ExpandQuery and ColTor.
    \item \textbf{Layout-aware RowSel optimizations:} 
    A transposed-layout RowSel with fine-grained pipelining that resolves the conflict between NTT-driven data layouts and tiled GEMMs.
    \item \textbf{Scalable multi-GPU execution:} 
    We design multi-GPU execution strategies to scale both query throughput and \DB capacity via \DB sharding and high-bandwidth inter-GPU communication, enabling efficient PIR execution.
    \item \textbf{State-of-the-art performance:} 
    Evaluation showing up to 297.2$\times$ higher throughput than PIRonGPU~\cite{github-PIRonGPU}, the prior state-of-the-art open-source implementation.
\end{itemize}
\begin{figure}[t]
    \centering
    \includegraphics[width=0.98\columnwidth]{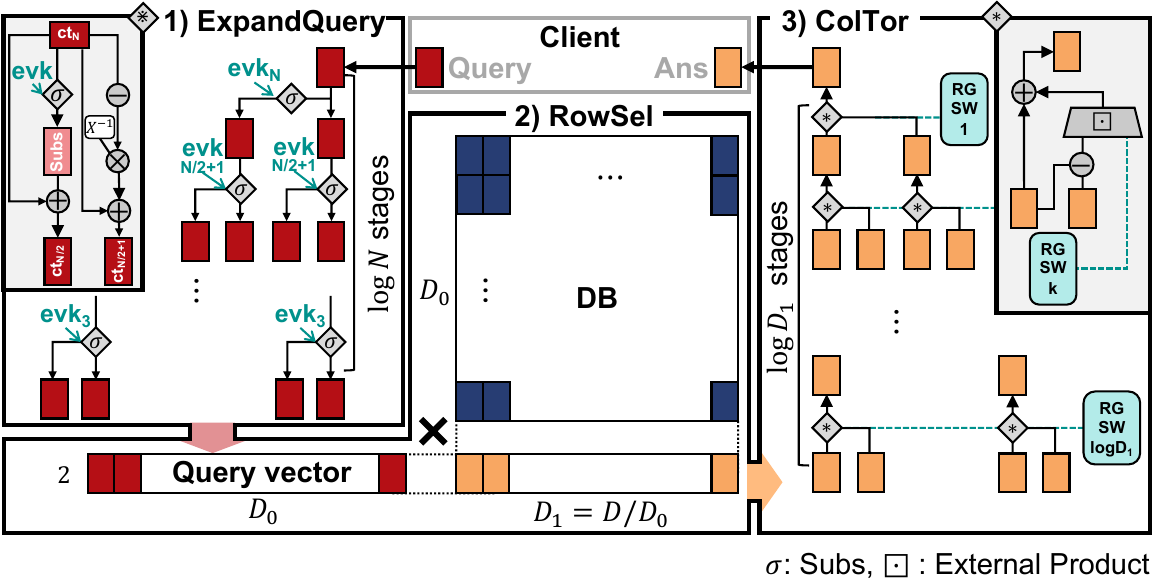}
    \Description{}
    \vspace{-0.08in}
    \caption{OnionPIRv2 computation process.}
    \label{fig:computation_process}
    \vspace{-0.04in}
\end{figure}

\section{Construction \& Bottlenecks of PIR}
\label{sec:background}

\subsection{Private Information Retrieval (PIR)}
\label{sec:background:pir}

Consider a client retrieving a record from a large unencrypted database (\DB) with $D$ records, which is maintained by a server.
PIR enables the client to obtain the $i^*$-th record $\mathDB[i^*]$ without disclosing $i^*$.
Lattice-based homomorphic encryption (HE)~\cite{stoc-2009-gentry} is typically used to encrypt the client's query index $i^*$ into HE ciphertext(s), with which the server performs encrypted retrieval of $\mathDB[i^*]$.

We target the most general PIR use case, free of restrictive assumptions such as multiple non-colluding servers~\cite{jacm-1998-chorpir}, a single client sending bulk queries at once~\cite{sp-2023-vectorized, sp-2018-sealpir}, or an offline phase for pre-downloading large data (hints~\cite{usenixsec-2023-simplepir} or even the entire database~\cite{sp-2024-piano}).
Although such assumptions can improve protocol efficiency, they inherently limit the broad applicability of PIR.

We focus on the OnionPIR family~\cite{ccs-2021-onionpir, sp-2022-spiral, iacr-2025-onionpirv2}, which incurs asymptotically minimum $\mathcal{O}(\log D)$ communication\footnote{The communication cost cannot become lower as the index size itself is $\mathcal{O}(\log D)$. We ignore factors from cryptographic parameters, such as $N$, for asymptotic analysis throughout this paper because they do not change significantly.} per query.
Many other popular PIR protocols, such as ~\cite{sp-2018-sealpir, security-2021-mulpir, popets-2016-xpir, popets-2023-frodopir, usenixsec-2023-simplepir}, require $\mathcal{O}(\sqrt{D})$ communication.
Minimizing communication overhead is critical:
once an efficient GPU implementation substantially reduces the computational bottlenecks, communication cost emerges as the primary limiting factor for overall performance.
Nonetheless, given the structural similarities among PIR protocols, our approach remains applicable to other schemes, such as ~\cite{sp-2022-spiral, ccs-2024-respire, ccs-2024-kspir, iacr-2024-whispir, iacr-2025-pir-revisited, sp-2026-via}, with minimal or no modification.

\subsection{Three Phases of OnionPIRv2}
\label{sec:background:protocol}

We introduce the computational process of OnionPIRv2~\cite{iacr-2025-onionpirv2}, an optimized variant of the OnionPIR family, consisting of three phases (Fig.~\ref{fig:computation_process}): ExpandQuery, RowSel, and ColTor.
HE operations from the BFV~\cite{crypto-2012-bfv, iacr-2012-fv} and RGSW~\cite{crypto-2013-gsw} HE schemes are used in each phase.

\DB is organized as a $D_0\times D_1$ structure, where each entry (record) $\mathDB[i][j]$ is a plaintext polynomial.
In \S\ref{sec:background:ring:explanation}, we demonstrate that the number-theoretic transform (NTT) maps a polynomial to a length-$4N$ (e.g., $N=2^{12}$) vector.
Within this transformed domain, polynomial addition and multiplication are simplified to point-wise addition and multiplication between the resulting vectors.

Initially, the client transmits a single BFV ciphertext (simply, \ciphertext) as a query to the server.
A BFV ciphertext consists of a pair of polynomials (i.e., two length-$4N$ vectors).
This query \ciphertext encapsulates the information for both the row index $i^*$ and the column index $j^*$ in a condensed format to minimize the communication overhead.

\noindent
\textbf{ExpandQuery:}
The server expands the query \ciphertext into up to $N$ \ciphertexts required for the following phases.
This is carried out through a binary-tree-shaped expansion process, where each tree node produces two output \ciphertexts from an input \ciphertext.
Each node performs a homomorphic substitution (\subs), mainly composed of NTT and digit decomposition ($\mathrm{Dcp}$) operations, explained in \S\ref{sec:background:dcp:explanation}.

We define each depth of the expansion tree as a stage of ExpandQuery.
Each stage requires an evaluation key (\evk), which is additional client-provided data essential for the \subs operations.
ExpandQuery consists of up to $\log N$ stages in total.

\noindent
\textbf{RowSel:}
After ExpandQuery, the server processes encrypted row-selection (RowSel).
Among the output \ciphertexts from ExpandQuery, a subset of $D_0$ \ciphertexts forms a one-hot representation of the row index $i^*$.
In this subset, every \ciphertext encrypts zero except for the $i^*$-th \ciphertext, which encrypts one.
The server multiplies this \ciphertext subset with \DB to extract the $i^*$-th row, accessing the entire \DB in the process.

The RowSel computation reduces to $4N$ parallel point-wise GEMMs, where each GEMM multiplies a $2 \times D_0$ matrix (\ciphertext polynomial points) by a $D_0 \times D_1$ matrix (\DB polynomial points).
RowSel produces $D_1$ ciphertexts as output, which consist of $2 \times D_1$ polynomials, equivalent to $2 \times D_1$ vectors of length $4N$.

\noindent
\textbf{ColTor:}
%
Finally, the server performs a tournament-like encrypted column selection (ColTor).
ColTor is computationally nearly identical to the reverse of ExpandQuery, having a binary-tree-shaped flow with $\log D_1$ stages.
Each tree node receives two input \ciphertexts and produces a single \ciphertext, performing an external product ($\boxdot$)~\cite{jc-2020-tfhe}, mainly composed of NTT and $\mathrm{Dcp}$ operations.
For the external product operations, each stage requires an RGSW ciphertext ($\mathtt{rgsw}$), derived from another subset of the ExpandQuery output \ciphertexts via minor computations.
The final result is a \ciphertext encrypting $\mathDB[i^*][j^*]$.

\begin{figure}[t]
    \centering
    \includegraphics[width=0.75\columnwidth]{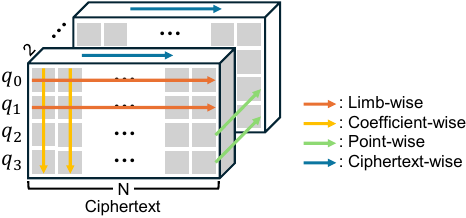}
    \vspace{-0.1in}
    \Description{}
    \caption{Four job partitioning strategies for PIR operations. Each cuboid represents a BFV \ciphertext with two polynomials, each having four limbs (rows) and $N$ coefficients (columns).}
    \label{fig:data_partition}
    \vspace{-0.1in}
\end{figure}

\subsection{Data Layout Conflict: NTT vs. GEMM}
\label{sec:limitation:data-layout}

A fundamental architectural challenge in GPU-based PIR is the data layout mismatch between phases.
It is convenient to formulate the parallel GEMMs in RowSel as $\mathbf{M}_\text{out}[p,m,n] = \sum_{k} \mathbf{M}_\text{in0}[p,m,k]\cdot \mathbf{M}_\text{in1}[p,k,n]$, where $p$, $m$, $n$, and $k$ dimensions have extents of $4N$, $2$, $D_1$, and $D_0$.
Standard HE GPU libraries, including PIRonGPU~\cite{github-PIRonGPU,iacr-2024-HEonGPU}, typically prioritize $p$-major layouts, ensuring the $4N$ points of each polynomial are contiguously allocated.
This arrangement is optimized to enable coalesced memory accesses for core polynomial operations and, in particular, NTT.
However, high-performance GEMM kernels require data to be contiguous along the $m$ (polynomial batch), $n$ (output), or $k$ (reduction) dimensions to enable efficient tiling and memory resource use~\cite{cutlass-blog}.

\subsection{Job Partitioning Conflicts in Polynomial Operations}
\label{sec:background:ring}

\subsubsection{RNS and NTT representation}
\label{sec:background:ring:explanation}

Each polynomial in OnionPIR is of degree $N-1$ with coefficients in the range $[-\frac{Q-1}{2}, \frac{Q-1}{2}]$ for a large modulus $Q$ (e.g., $Q\simeq 2^{108}$).
To avoid high-precision arithmetic for large $Q$, the residue number system (RNS) decomposes $Q$ into multiple RNS primes~\cite{rsa-2019-rns-bfv,crypto-2012-aes}.
We specifically use four 32-bit primes to set $Q=q_0q_1q_2q_3$ throughout this paper.
Each polynomial coefficient is reduced modulo each $q_i$ to produce a length-$N$ 32-bit vector, referred to as a \emph{limb}.
The resulting four limbs comprise the $4N$ points of the polynomial.
This can also be structured as a $4 \times N$ matrix, where each row corresponds to a limb and each column to a coefficient (see Fig.~\ref{fig:data_partition}).

With RNS, adding two polynomials is equivalent to performing point-wise additions modulo $q_i$ between two length-$4N$ vectors.
By contrast, multiplying two polynomials requires limb-wise (negacyclic) convolutions, which are facilitated by the number-theoretic transform (NTT).
NTT is a variant of the Fourier transform that converts limb-wise convolutions ($*$) into point-wise multiplications ($\odot$); i.e., $\mathrm{NTT}(a * b) = \mathrm{NTT}(a) \odot \mathrm{NTT}(b)$.
Besides the GEMMs for RowSel, NTT is the most dominant primitive operation in PIR~\cite{hpca-2026-ive}.

\subsubsection{Job partitioning conflicts}
\label{sec:background:ring:bottleneck}

Polynomial operations require different partitioning strategies (see Fig.~\ref{fig:data_partition}) to accommodate their unique data dependencies.
For instance, limb-wise partitioning is adequate for NTT; as each limb is transformed independently, polynomial limbs can be distributed across different processors (SMs in NVIDIA GPUs, \S\ref{sec:background:gpu}) without any inter-processor communication.
In contrast, RNS-related operations such as $\mathrm{Dcp}$ operate per coefficient, making coefficient-wise partitioning the preferred choice.

However, exploiting such fine-grained parallelism within each polynomial introduces a significant bottleneck: switching between limb-wise and coefficient-wise partitioning requires inter-processor communication.
Because the data required for one partitioning scheme is distributed across different physical processors in the other, costly global synchronization and data transfers are required to transition between these two partitioning strategies.

Meanwhile, ciphertext-wise partitioning, adopted in the block-based batching strategy of \cite{cal-2024-tfhe}, can support both NTT and $\mathrm{Dcp}$.
However, all these partitioning strategies conflict with point-wise operations, including the parallel GEMMs in RowSel.

\subsection{Digit Decomposition \& Cache-Capacity Wall}
\label{sec:background:dcp}

\subsubsection{$\mathrm{Dcp}$ for \subs and external product}
\label{sec:background:dcp:explanation}

The core computations in ExpandQuery and ColTor---homomorphic substitution (\subs) and external product---follow similar workflows centered on digit decomposition ($\mathrm{Dcp}$). $\mathrm{Dcp}$ extracts digits from each polynomial coefficient using a base $z$ with $z^\ell > Q$ (e.g., $z=2^{22}$, $\ell=5$ for $Q\simeq2^{108}$), producing $\ell$ polynomials whose coefficients lie in $[-\frac{z}{2}, \frac{z}{2} + 1]$\footnote{
ExpandQuery generates $D_0 + \log_2(D_1)\cdot \ell$ \ciphertexts, which is much smaller than $N$ under our parameter settings following OnionPIRv2~\cite{iacr-2025-onionpirv2}. Among these, the $D_0$ subset is used for RowSel, while the $\log_2(D_1)\cdot \ell$ subset is used for $\mathtt{rgsw}$.}.

RNS and NTT complicate the computation process as follows:
\begin{enumerate}[leftmargin=*]
    \item \emph{Inverse NTT} is applied to the input NTT-domain polynomial.
    \item \emph{Dcp} now includes \emph{RNS reconstruction}, which recovers the large-coefficient polynomial from its four limbs. The actual decomposition follows, generating $\ell$ digit polynomials. The entire $\mathrm{Dcp}$ is carried out through coefficient-wise computations.
    \item \emph{Forward NTT} is performed on the $\ell$ polynomials for subsequent computations in the NTT domain.
\end{enumerate}

For a ciphertext comprising two polynomials (e.g., $\mathtt{ct}=(a,b)$), $\mathtt{Subs}$ performs $\mathrm{Dcp}$ only on $a$, while the external product performs $\mathrm{Dcp}$ on both $a$ and $b$.
In addition to $\mathrm{Dcp}$, both operations execute polynomial multiply-and-accumulate tasks between the $\ell$ (or $2\ell$) decomposed polynomials and an $\mathtt{evk}$ (or an $\mathtt{rgsw}$).
This accumulation ultimately results in two polynomials.

\subsubsection{Transient working-set spikes}
\label{sec:background:dcp:bottleneck}

$\mathrm{Dcp}$ substantially increases the memory capacity requirements during ExpandQuery/ColTor due to the $\ell\times$ expansion in the number of polynomials.
However, because the $\ell$ output polynomials are immediately consumed by the subsequent accumulation, they create transient spikes in the working set size rather than a permanent memory footprint.

While an effective caching strategy could prevent these transient spikes from becoming a performance bottleneck, the limited cache capacity in GPUs restricts cache reuse.
This limitation eventually necessitates DRAM transfers to store and load the $\ell$ polynomials, particularly as the number of \ciphertexts increases at deeper tree depths.

Consequently, we identify a \textbf{cache-capacity wall} in ExpandQuery and ColTor, where performance becomes constrained by excessive DRAM transfers as the transient working set size exceeds the last-level (L2) cache capacity of the GPU.

\subsection{Multi-GPU Scaling}
\label{sec:background:gpu}

\textbf{GPU execution model:}
A GPU comprises hundreds of streaming multiprocessors (SMs) that execute threads at warp (32 threads per warp) granularity under a single-instruction, multiple-thread (SIMT) model.
Threads are grouped into thread blocks, each scheduled on a single SM.
A grid of thread blocks forms a GPU kernel.
To hide execution latency, SMs context-switch among active warps; thus, performance depends on occupancy, defined as the ratio of active warps to the hardware maximum.
High occupancy ensures the SM remains productive while some warps stall on memory or arithmetic operations.

The GPU memory hierarchy spans fast on-chip storage (registers, shared memory, and L1/L2 caches) and off-chip DRAM. 
In multi-GPU configurations, devices communicate via PCIe or NVLink.
Due to the severe disparity between internal DRAM bandwidth and the significantly lower bandwidth of inter-GPU interconnects, data must be explicitly managed to minimize communication bottlenecks while also balancing aggregate throughput.

\noindent
\textbf{Multi-GPU scaling challenges:}
To improve the applicability of GPU-based PIR for practical large-scale \DB, we can utilize multiple GPUs to shard the $\mathtt{DB}$, gaining a proportional increase in both DRAM capacity and bandwidth.
As each PIR query necessitates reading the entire $\mathtt{DB}$, maintaining the dataset within the aggregate GPU DRAM capacity is essential for performance.

While handling RowSel through distributed GEMMs across a sharded $\mathtt{DB}$ is straightforward, orchestrating distributed processing for ExpandQuery and ColTor remains challenging.
Without careful management, the communication and synchronization overheads, induced by sharing expanded \ciphertexts or merging partial tournament results, can negate the throughput gains of multi-GPU execution.
To the best of our knowledge, no prior work has explored these multi-GPU coordination challenges in PIR.

\section{Impact Analysis of Multi-Client Batching}
\label{sec:limitation}

\begin{figure}[t]
    \centering
    \includegraphics[width=0.99\columnwidth]{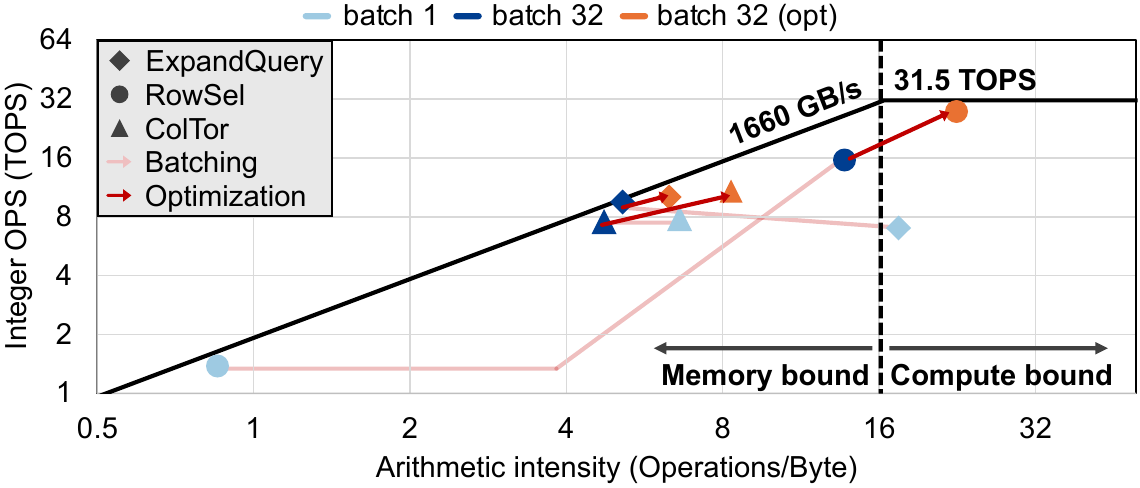}
    \Description{}
    \vspace{-0.08in}
    \caption{Roofline analysis of each PIR phase on an RTX 5090 with a 2\,GB \DB. Computational throughput is measured in terms of 32-bit integer multiply-and-add (IMAD) throughput.}
    \label{fig:roofline}
    \vspace{-0.03in}
\end{figure}

While most PIR studies focus on single-query latency, multi-client batching can fundamentally alter the performance characteristics of PIR workloads.
In particular, the RowSel phase requires scanning the entire \DB, making memory bandwidth a dominant bottleneck.
Multi-client batching amortizes the \DB access cost across multiple queries, enabling more practical deployment of PIR services~\cite{hpca-2026-ive}.
We perform a detailed throughput and memory analysis on an NVIDIA RTX~5090, which provides high integer throughput (31.5 TOPS) and DRAM bandwidth (1660\,GB/s).

\subsection{Bottleneck Shift Due to Batching}
\label{sec:limitation:batching}

First, we conducted a roofline analysis of the three PIR phases.
We used the baseline ``operation-level'' kernels, which will be described in \S\ref{sec:opt-1:baseline}.
Our analysis reveals mixed benefits of multi-client batching across different phases, as shown in Fig.~\ref{fig:roofline}.

\subsubsection{RowSel: From memory-bound to layout-limited}

Multi-client batching fundamentally changes the performance regime of RowSel, exposing data layout as a first-order bottleneck.
Without batching, RowSel exhibits extremely low arithmetic intensity due to the
small $m$ dimension ($m=2$, \S\ref{sec:limitation:data-layout}), rendering the operation strongly constrained by memory bandwidth.
In this regime, as long as the available DRAM bandwidth is fully utilized,
the specific data layout has a limited impact on performance.

Batching increases the $m$ dimension to $2 \cdot \text{batch}$, raising arithmetic intensity from 0.9 to 13.8~Ops/Byte at batch size 32 and moving RowSel close to the RTX~5090 roofline ridge point (Fig.~\ref{fig:roofline}). 
This shift transitions RowSel toward a compute-bound regime.
However, in this regime, performance becomes highly sensitive to data layout. Despite higher arithmetic intensity, RowSel achieves 2.0$\times$ lower throughput than the theoretical peak, indicating that data layout conflicts, negligible in the memory-bound regime, now dominate performance by limiting effective GPU utilization.

\subsubsection{ExpandQuery and ColTor: Batching-induced memory capacity pressure}

In contrast, multi-client batching intensifies memory pressure in ExpandQuery and ColTor.
ExpandQuery and ColTor primarily operate on client-specific data, such as \ciphertexts and \evks.
Thus, they do not benefit from inter-batch data sharing.
Instead, larger batch sizes inflate the working set, degrading cache locality and causing frequent spills from the 96\,MB L2 cache of the RTX~5090 to off-chip DRAM.
Consequently, increases in off-chip DRAM traffic reduce the arithmetic intensity of ExpandQuery and ColTor under batching, pushing these operations deeper into the memory-bound region rather than alleviating memory pressure.

\begin{figure}[t]
    \centering
    \includegraphics[width=0.99\columnwidth]{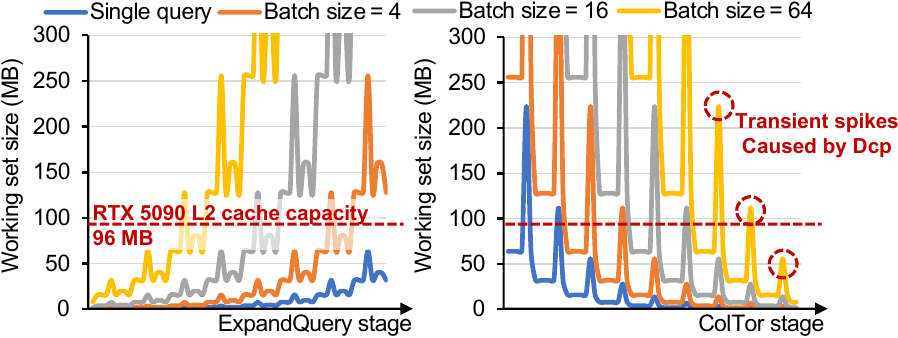}
    \Description{}
    \vspace{-0.05in}
    \caption{Operation-wise working set variation across ExpandQuery and ColTor phases under different batch sizes.}
    \label{fig:working-set}
    \vspace{-0.05in}
\end{figure}

\subsection{Batching-Amplified Working-Set Spikes}
\label{sec:limitation:working-set}
A deeper investigation into the memory behavior of ExpandQuery and ColTor (see Fig.~\ref{fig:working-set}) reveals that multi-client batching magnifies the transient working-set spikes caused by $\mathrm{Dcp}$.
Even at shallow tree depths, these spikes increasingly approach the cache-capacity wall with batching.
While such spikes already exist in single-query execution, batching causes numerous such spikes to coexist across queries, rapidly amplifying memory capacity demand and turning a transient effect into a persistent bottleneck.

In ExpandQuery, each stage operates only on the $a$ polynomial of each \ciphertext during \subs, yielding a 64\,KB memory footprint ($4\times N\times 32\textit{bit}$).
While this memory footprint remains within the L2 cache under single-query execution, the number of active \ciphertexts doubles as the tree depth increases; under multi-client batching, this growth further compounds across queries, quickly pushing the aggregate working set beyond the 96\,MB L2 cache capacity of the RTX~5090.

This effect is further amplified in ColTor, where both $a$ and $b$ polynomials are processed during the external product, doubling the footprint to 128\,KB.
For example, at the first ColTor stage with 256 active \ciphertexts and $\ell=5$,
a batch size of 32 already incurs a transient working set of approximately
$256 \times 5 \times 128\,\text{KB} \times 32 \approx 5$\,GB.
This footprint far exceeds on-chip cache capacity,
illustrating how batching rapidly accelerates the onset of the cache-capacity wall.

Consequently, batching transforms localized, short-lived memory expansions into sustained pressure on the memory hierarchy.
Under an operation-level kernel, these batching-amplified intermediates must be repeatedly written to and reloaded from DRAM, causing cache pollution and increased off-chip memory traffic.

\subsection{Design Implications}

Our analysis demonstrates that multi-client batching impacts PIR phases in different ways.
For RowSel, batching raises arithmetic intensity but leaves a gap to peak GPU throughput due to data layout inefficiencies.
For ExpandQuery and ColTor, batching instead amplifies working-set growth and intensifies the cache-capacity wall problem, making memory traffic the dominant bottleneck.

This discrepancy suggests that a uniform GPU execution strategy is suboptimal for batched PIR.
\emph{Efficient designs must instead adapt execution and data layout to the dominant bottleneck of each phase, motivating stage-aware and layout-aware techniques introduced in the following sections.}

\section{Stage-Aware Hybrid Execution for ExpandQuery and ColTor}
\label{sec:opt-1}
The analysis in \S\ref{sec:limitation} shows that, under multi-client batching, ExpandQuery and ColTor become dominated by memory behavior rather than computation. 
Consequently, kernel granularity is a critical design choice, as intermediate data movement can significantly inflate off-chip traffic. 
Motivated by this, we analyze operation-level and stage-level kernels which fuse operations within a stage for ExpandQuery and ColTor in the following subsections.
\subsection{Operation-Level vs. Stage-Level Kernels}
\label{sec:opt-1:analysis_1}

\subsubsection{Baseline operation-level kernels}
\label{sec:opt-1:baseline}

We build our baseline implementation based on PIRonGPU~\cite{github-PIRonGPU, iacr-2024-HEonGPU}, which exploits fine-grained job partitioning adequate for each polynomial operation.
We refer to the resulting kernel implementations as \emph{operation-level} kernels.
Individual polynomial operations are implemented as separate kernels, each employing distinct job partitioning strategies to assign tasks to thread blocks.
For example, limb-wise partitioning is used for NTT, coefficient-wise for $\mathrm{Dcp}$, and point-wise for polynomial addition and multiplication.
Following IVE~\cite{hpca-2026-ive}, our GPU baseline adopts multi-client batching.

\subsubsection{Stage-level kernels}
\label{sec:opt-1:stage-level}

In addition to the baseline operation-level kernels, we introduce \emph{stage-level} kernels that process an entire stage of ExpandQuery/ColTor within a single kernel invocation.
A stage-level kernel fuses all operation-level kernels within a stage of ExpandQuery/ColTor, such as NTT, $\mathrm{Dcp}$, and multiplications with \evks or RGSW ciphertexts, into a single kernel.

Through kernel fusion, stage-level kernels retain intermediate results---specifically the transient polynomials produced during $\mathrm{Dcp}$---within SM registers or shared memory.
This localized data handling effectively avoids hitting the cache-capacity wall, alleviating the performance bottlenecks otherwise caused by the limited L2 cache capacity and working-set spikes.

However, stage-level kernels reduce parallelism, which can adversely impact performance.
Operation-level kernels exploit not only ciphertext-wise partitioning but also limb-wise and coefficient-wise partitioning to distribute work across thread blocks.
However, fusing these operation-level kernels into a single stage-level kernel introduces interleaved partitioning patterns that create all-to-all data dependencies within each \ciphertext, requiring each thread block to process an entire \ciphertext (i.e., a single tree node).
As a result, stage-level kernels substantially reduce the number of thread blocks, which can lower occupancy and degrade performance.

Multi-client batching increases the number of \ciphertexts and makes stage-level kernels increasingly viable, as ciphertext-wise partitioning alone can provide sufficient parallelism to fully utilize GPU resources.
The ExpandQuery/ColTor tree structure exposes two sources of ciphertext-wise data parallelism: (1) parallel processing of multiple nodes at the same stage (tree depth) and (2) parallelism across multiple \ciphertexts introduced by multi-client batching.
While only dozens of thread blocks can be populated at shallow tree depths with stage-level kernels, up to tens of thousands of thread blocks can be populated for deeper tree depths.

\begin{figure*}[t]
    \centering
    \subfloat[DRAM transaction (ExpandQuery)]{\includegraphics[width=0.45\linewidth]{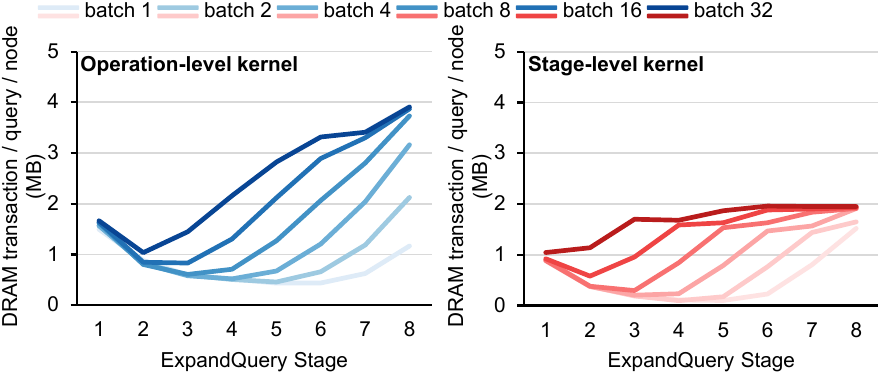}\label{fig:analysis:data-acceess-MK}\Description{}}
    \hspace{0.5cm}
    \subfloat[DRAM transaction (ColTor)]{\includegraphics[width=0.49\linewidth]{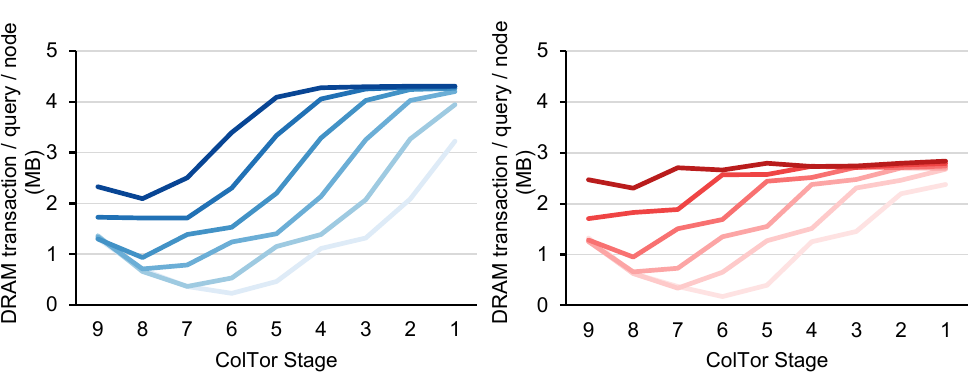}\label{fig:analysis:data-access-SK}\Description{}}
    \hfill
    \subfloat[Execution time (ExpandQuery)]{\includegraphics[width=0.45\linewidth]{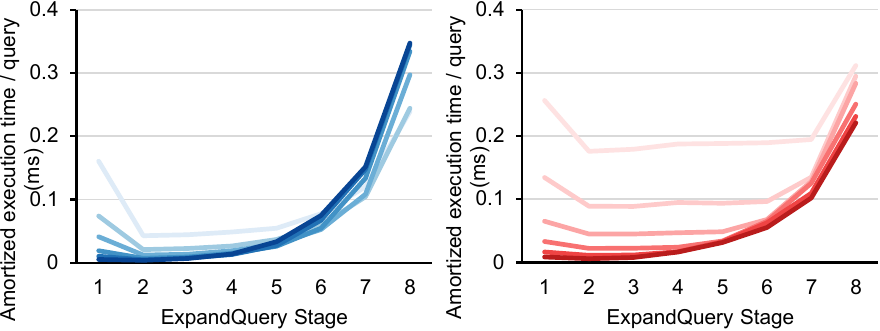}\label{fig:analysis:exec-time-MK}\Description{}}
    \hspace{0.5cm}
    \subfloat[Execution time (ColTor)]{\includegraphics[width=0.49\linewidth]{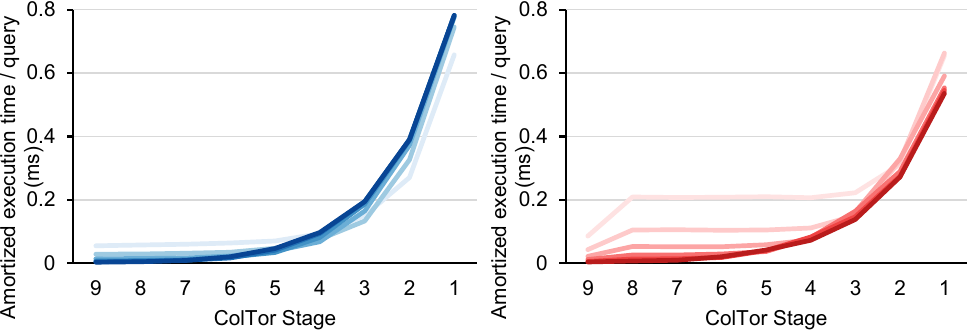}\label{fig:analysis:exec-time-SK}\Description{}}
    \vspace{-0.05in}
    \caption{Stage-wise DRAM transaction and execution time of operation- and stage-level kernels under various batch sizes. (a, b) DRAM transactions per query per node and (c, d) Amortized execution time per query for ExpandQuery/ColTor.} 
    \label{fig:analysis}
    \vspace{-0.05in}
\end{figure*}

\subsection{Stage-Wise Comparison of Kernel Behaviors}
\label{sec:opt-1:profile}

To quantitatively analyze the trade-offs between operation-level and stage-level kernel designs, we profile DRAM traffic and execution time for each stage.
We focus primarily on the memory behavior as both ExpandQuery and ColTor are constrained by DRAM bandwidth under batching (see Fig.~\ref{fig:roofline}).
Our analysis uses NVIDIA Nsight Compute~\cite{nsight-compute} on an NVIDIA RTX~5090 GPU.

\subsubsection{DRAM traffic trends}

Fig.~\ref{fig:analysis:data-acceess-MK} and Fig.~\ref{fig:analysis:data-access-SK} show that DRAM traffic exhibits strong stage dependence under batching, and that the two kernel designs respond differently to working-set growth.
For early stages in ExpandQuery and later stages in ColTor, where the number of \ciphertexts is small, the working set fits within the L2 cache (see Fig.~\ref{fig:working-set}), resulting in low DRAM traffic for both operation-level and stage-level kernel designs.
In this cache-resident regime, DRAM transactions per query per node further decrease as the number of tree nodes increases (rightward in Fig.~\ref{fig:analysis:data-acceess-MK} and Fig.~\ref{fig:analysis:data-access-SK}), as \evks and RGSW ciphertexts are shared across nodes.

As the batch size increases, the stage working set eventually exceeds the L2 cache capacity, even for stages with few nodes.
Beyond this point, the operation-level kernel design incurs sharp increases in DRAM traffic, as intermediate results are repeatedly written to and read from DRAM between kernels.
In contrast, the stage-level kernel design substantially reduces DRAM transactions (up to 2.01$\times$) by consuming intermediate data within a single kernel invocation and exploiting temporal locality.
This divergence is most pronounced in later ExpandQuery and earlier ColTor stages, where a large number of nodes must be processed.

\subsubsection{Execution time and occupancy}

The execution-time trends in Fig.~\ref{fig:analysis:exec-time-MK} and Fig.~\ref{fig:analysis:exec-time-SK} reflect the interplay between memory traffic and attainable degree of parallelism under the two kernel designs.
Stages whose working sets exceed the L2 cache capacity become bottlenecked by DRAM bandwidth.
In this regime, execution times closely follow the DRAM traffic trends, and the stage-level design consistently outperforms the operation-level design by minimizing DRAM accesses, achieving up to 1.57$\times$ lower amortized execution time per query.
Under this DRAM bandwidth bottleneck, increasing the batch size for operation-level kernels can even adversely impact performance by further enlarging the working set.

In contrast, in the early stages of ExpandQuery and the later stages of ColTor, with the limited number of active \ciphertexts, occupancy becomes the dominant performance factor.
As the operation-level kernel launches substantially more thread blocks, it achieves higher SM occupancy and lower execution time than the stage-level kernel.

\subsection{Memory Bandwidth Scaling Benchmark}
\label{sec:opt-1:microbench}

\begin{figure}[t]
    \centering
    \includegraphics[width=0.9865\columnwidth]{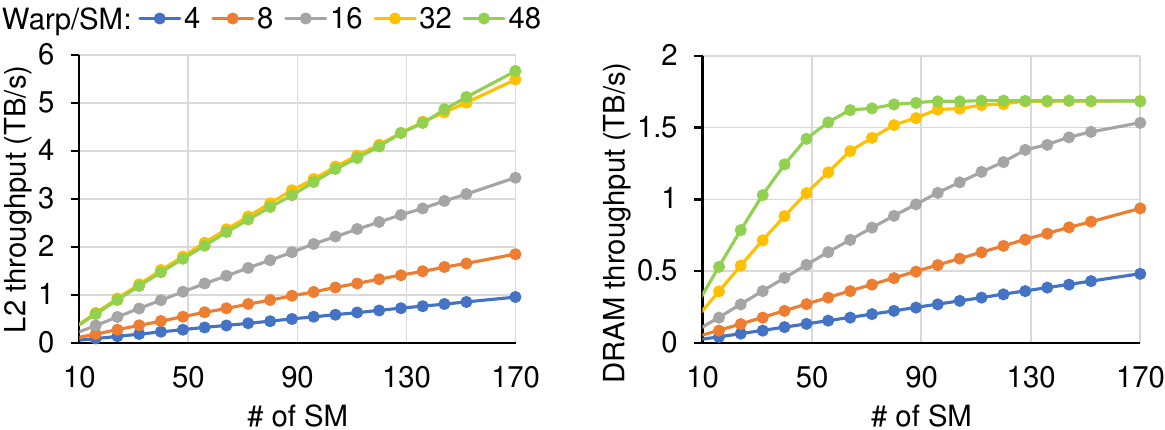}
    \vspace{-0.05in}
    \Description{}
    \caption{Achieved throughput under a stream-like benchmark. L2 and DRAM throughput are measured while varying the number of active SMs and warps per SM. L2 uses a working set fitting in the 96\,MB cache, whereas DRAM uses a 2\,GB working set exceeding cache capacity.}
    \label{fig:BW_vs_Occupancy}
    \vspace{-0.05in}
\end{figure}

Occupancy, which is determined by the number of thread blocks, not only affects the computational throughput but also determines the attainable global memory (L2 cache and DRAM) bandwidth.
We conduct stream-like benchmarks that measure sustained L2 and DRAM bandwidth while varying the number of active SMs and the number of warps per SM (see Fig.~\ref{fig:BW_vs_Occupancy}).

The results show that L2 bandwidth exhibits a strong dependence on both inter-SM and intra-SM parallelism (see the left side of Fig.~\ref{fig:BW_vs_Occupancy}).
With a small number of active SMs, the achieved L2 throughput remains far below the hardware peak, even though all data are served from on-chip cache.
As the number of active SMs increases, L2 bandwidth scales almost linearly.
Maximizing L2 bandwidth also requires launching a large number of warps per SM, as evidenced by bandwidth increasing steadily up to 32 warps per SM.

In contrast, DRAM bandwidth saturates at much lower occupancy, which highlights a fundamental asymmetry between on-chip and off-chip memory systems (see the right side of Fig.~\ref{fig:BW_vs_Occupancy}). 
When using all the SMs, doubling the number of per-SM warps from 16 to 32 results in a mere 9.7\% improvement in achieved DRAM bandwidth.
While exploiting L2 bandwidth requires high occupancy and abundant parallelism, DRAM-bound execution is far less sensitive to occupancy once bandwidth saturation is reached.
Notably, these trends are independent of PIR-specific computation and reflect fundamental properties of the GPU memory hierarchy.

\subsection{Stage-Aware Hybrid Kernel Execution}
\label{sec:opt-1:hybrid_kernel}

Building on the results from \S\ref{sec:opt-1:profile} and \S\ref{sec:opt-1:microbench}, we propose a stage-aware hybrid execution strategy that selects the kernel execution model based on the dominant bottleneck at each stage.
Specifically, we use operation-level kernels when a stage processes a small number of \ciphertexts such that the working set mostly remains L2-resident, and stage-level kernels otherwise.

We use operation-level kernels when the working set of a stage mostly remains L2-resident.
Under such circumstances, performance is primarily determined by the ability to exploit on-chip memory bandwidth.
As discussed in \S\ref{sec:opt-1:microbench}, effective utilization of L2 bandwidth requires relatively high occupancy.
Operation-level kernels are preferred because they exploit additional intra-ciphertext parallelism, populating more thread blocks.

Once the working set exceeds the L2 cache capacity, we switch to stage-level kernels to mitigate the DRAM bandwidth bottleneck.
In this regime, the number of \ciphertexts is sufficient to fully utilize the SMs, achieving occupancy beyond the saturation point (16 warps per SM) using ciphertext-wise partitioning alone.
As higher occupancy does not yield additional DRAM bandwidth, reducing DRAM traffic through stage-level kernels becomes the superior option.

We define the point at which the stage working set exceeds L2 capacity as the transition boundary between the two execution models.
By switching kernel design at this boundary, the hybrid design maximizes occupancy when on-chip reuse is effective and minimizes DRAM traffic when reuse fails, achieving robust performance across all stages of the PIR computation. This boundary can be determined statically from batch size and tree depth, and does not require runtime profiling.

\section{Layout-Aware RowSel Optimizations}
\label{sec:opt-2}

Although multi-client batching increases RowSel's arithmetic intensity and moves it toward a more balanced point in the roofline plot (see Fig.~\ref{fig:roofline}), the baseline batched RowSel implementation, based on observations from IVE~\cite{hpca-2026-ive}, falls short of fully exploiting the GPU’s peak computational throughput.

We identify that this performance gap does not stem from insufficient computational resources, but from a structural mismatch between the NTT-oriented data layout and the data access pattern required by batched RowSel, which is implemented as matrix--matrix multiplications (GEMMs).

\subsection{Tiled RowSel GEMMs with Layout Conflicts}
\label{sec:limitation:analysis_2}

\subsubsection{Tiled GEMMs for RowSel}

With batching, RowSel is transformed into $4N$ parallel GEMMs, each with a $(2\cdot\mathsf{batch})\times D_0$ matrix and a $D_0 \times D_1$ matrix.
As GEMM with sufficiently large dimensions is compute-bound, large $\mathsf{batch}$ sizes, such as 32 or 64, can render RowSel compute-bound.

We apply well-known tiling methods~\cite{cutlass-blog} to the parallel GEMMs.
For a regular GEMM multiplying an $m\times k$ matrix with an $n\times k$ matrix, work is distributed across thread blocks using tiles parameterized by $\mathrm{B}_m$, $\mathrm{B}_n$, and $\mathrm{B}_k$.
Each thread block computes a $\mathrm{B}_m\times \mathrm{B}_n$ output tile and iterates over the reduction dimension $k$ in steps of $\mathrm{B}_k$, loading input tiles into shared memory (requiring $B_m \cdot B_k + B_n \cdot B_k$ capacity) and accumulating partial results in registers.
The final output is written back to global memory.

Selecting $\mathrm{B}_m$, $\mathrm{B}_n$, and $\mathrm{B}_k$ entails a trade-off between global memory traffic and on-chip resource usage, necessitating careful adjustment~\cite{iccd-2023-cutailer,ppopp-2019-gemm,icpads-2010-cachegemm}.
Larger tiles reduce global memory traffic and DRAM pressure, but increase shared memory and register usage, potentially lowering occupancy and performance.

For RowSel, we have an additional $p$ dimension for $4N$ points of each polynomial.
General HE implementations prefer $p$-major data layouts to allow contiguous access for polynomial operations and, in particular, NTT.
However, this NTT-driven $p$-major data layout poses a dilemma for implementing the parallel GEMMs with tiling.
Enforcing memory coalescing under this layout requires each thread block to process multiple GEMMs concurrently, increasing on-chip memory requirements and reducing occupancy.
By contrast, avoiding this concurrency leads to non-coalesced accesses and significantly lower effective global memory bandwidth.

\begin{figure}[t]
    \centering
    \subfloat[Full RowSel GEMMs]{\includegraphics[width=0.33\columnwidth]{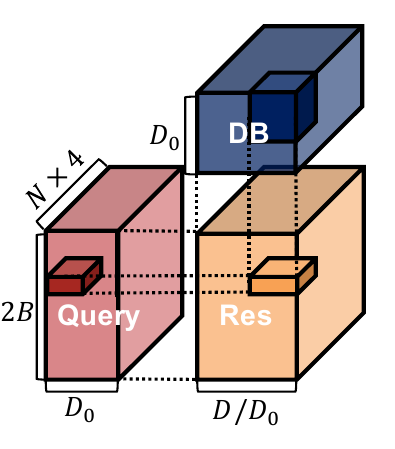}\label{fig:opt:data-layout-data}\Description{}}
    \subfloat[Tiling (baseline)]{\includegraphics[width=0.33\columnwidth]{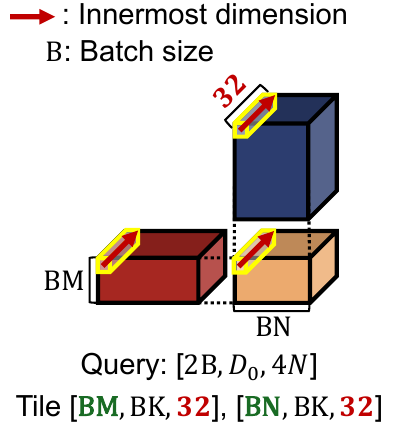}\label{fig:opt:data-layout-baseline}\Description{}}
    \subfloat[Tiling (transposed)]{\includegraphics[width=0.33\columnwidth]{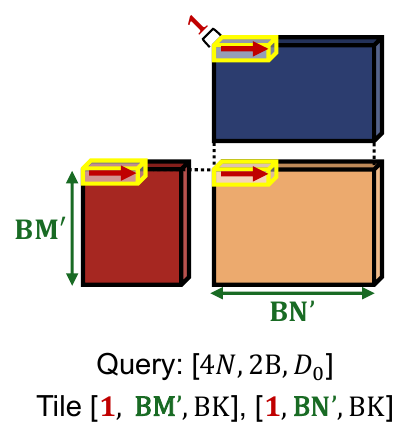}\label{fig:opt:data-layout-transpose}\Description{}}
    \caption{(a) Required GEMMs for RowSel and (b)(c) two thread-block tiling strategies for these GEMMs. The baseline tiling (b) uses a $p$-major data layout, whereas the optimized transposed-layout tiling (c) adopts a $k$-major data layout.}
    \label{fig:data-layout}
\end{figure}

\subsubsection{Performance degradation in $p$-major RowSel}
\label{sec:limitation:analysis_2:degradation}

We start from a baseline $p$-major RowSel implementation, where we perform GEMMs between a tensor of $\mathsf{Shape}(2\cdot\mathsf{batch},D_0,4N)$ corresponding to the input \ciphertexts and a tensor of $\mathsf{Shape}(D_1,D_0, 4N)$ corresponding to \DB.
Here, the sizes of $p$, $m$, $n$, and $k$ dimensions are $4N$, $2\cdot\mathsf{batch}$, $D_1$, and $D_0$, respectively.
For tiling, besides $\mathrm{B}_m$, $\mathrm{B}_n$, and $\mathrm{B}_k$, the baseline implementation introduces $\mathrm{B}_p$ to satisfy memory coalescing requirements.
Each thread block handles $\mathrm{B}_p$ parallel GEMMs.

The baseline RowSel kernel fails to exploit the full computational throughput of a GPU.
The RTX~5090 has a peak 32-bit integer multiply-add (IMAD) throughput of 31.5 TOPS.
However, as shown in Fig.~\ref{fig:roofline}, the baseline RowSel implementation for a 2\,GB \DB with a batch size of 32 achieves only  15.6 TOPS of IMAD throughput.
This occurs because the baseline RowSel is limited by L2 cache throughput (54.08\%) rather than being compute-bound as predicted by the theoretical analysis, and its achievable occupancy is capped at 33\% due to excessive register and shared memory footprint.

\subsection{Optimized Transposed-Layout RowSel}

\subsubsection{Transposed-layout GEMMs}

We explicitly transpose the data to enable $k$-major or $m$/$n$-major data layouts that are better suited for GEMMs.
For \DB, the layout can be chosen freely as the transposition can be performed in advance, before receiving any query.
The \ciphertexts can be reorganized into a $\mathsf{Shape}(4N, D_0, 2 \cdot \mathsf{batch})$ tensor during runtime (after ExpandQuery) to support an $m$-major layout.

With the layout restriction removed, we explore tiling configurations for the transposed-layout GEMMs.
For our exemplar 2\,GB \DB with a batch size of 32, we use $(\mathrm{B}_m,\,\mathrm{B}_n,\,\mathrm{B}_k)=(64,\,64,\,32)$ for the transposed layouts, whereas we are forced to use $(16,\,16,\,8)$ for the baseline with $\mathrm{B}_p = 32$ for $p$-dimension coalescing.
The additional $\mathrm{B}_p$ necessitates reducing $\mathrm{B}_m$, $\mathrm{B}_n$, and $\mathrm{B}_k$ to fit shared memory and register constraints.
Even so, the baseline configuration utilizes 2$\times$ more shared memory ($\propto(\mathrm{B}_m\cdot \mathrm{B}_k + \mathrm{B}_n \cdot \mathrm{B}_k)\cdot \mathrm{B}_p$), 2$\times$ more registers for accumulation ($\propto\mathrm{B}_m\cdot\mathrm{B}_n\cdot\mathrm{B}_p$), and 4$\times$ more global memory accesses ($\propto \frac{1}{B_m}$ when $B_m=B_n$).

Compared to the baseline (15.6 TOPS), the transposed-layout GEMMs reduce DRAM accesses and achieve significantly higher computational throughput (27.5 TOPS), as illustrated in Fig.~\ref{fig:roofline}.
Explicit transposition shifts RowSel into the compute-bound regime, achieving 86.88\% of the peak computational throughput.
Compared to the baseline, we achieve 1.56$\times$ higher memory throughput and a 1.5$\times$ increase in occupancy due to reduced per-block register and shared-memory footprint.
These results demonstrate that careful data layout reshaping is essential for sustaining high GEMM throughput under multi-client batching.

\subsubsection{$p$-dimension pipelining to reduce transpose overheads}

\begin{figure}
    \centering
    \includegraphics[width=0.97\columnwidth]{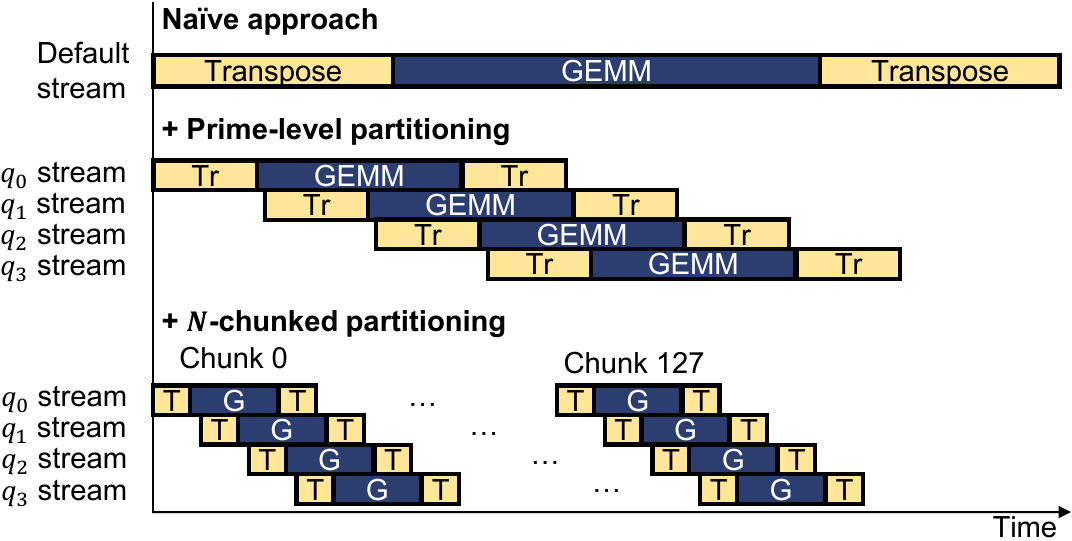}
    \caption{Gradually applying additional partitioning methods (+Prime-level partitioning, +N-chunked partitioning) for pipelined execution of our transposed-layout RowSel.}
    \label{fig:pipelining}\Description{}
\end{figure}

Launching two transpose kernels, one before RowSel and one after RowSel (na\"ive approach in Fig.~\ref{fig:pipelining}), introduces significant execution time overheads from loading and storing the input \ciphertexts from global memory.
For our 32-batch 2\,GB \DB example, input and output transpositions incur at least 6\,GB of memory traffic, which corresponds to 3.7\,ms of minimum latency.
Given that the transposed-layout GEMM kernel itself takes 10.9\,ms, the overhead accounts for a substantial fraction of the total RowSel execution time. 

Simply placing the kernels in different CUDA streams to hide latency does not lead to an automatic overlap between the kernels.
The RowSel GEMM kernel launches a very large grid along the $p$ dimension, corresponding to thousands of independent points per polynomial.
This saturates nearly all SM resources, leaving no scheduling headroom for concurrent execution of the transpose kernels.
Even when issued in separate streams, the excessive parallelism along the $p$-dimension prevents effective overlap, forcing the transpositions to be serialized behind the GEMMs.

To create scheduling headroom and enable overlap, we partition RowSel execution along the $p$-dimension at a fine granularity, decomposing both the RNS-prime and $N$ axis.
We first assign RowSel computations for different RNS primes $(q_0, q_1, q_2, q_3)$ to separate CUDA streams.
While this coarse-grained prime-level partitioning reduces transposition latency overheads, it introduces pipeline fill and drain overheads, as illustrated in the center diagram of Fig.~\ref{fig:pipelining}.
Therefore, we further split the $N=2^{12}$ points in each limb into multiple chunks and launch a separate kernel for each chunk, enabling a finer overlap between the kernels.
By combining prime-level stream partitioning with $N$-chunked launches, we create overlap opportunities between the input/output transpositions and the GEMMs across streams ($N$-chunked partitioning in Fig.~\ref{fig:pipelining}).
To mitigate the increased kernel launch overheads from this fine-grained partitioning, we utilize CUDA Graphs~\cite{cuda-programming-guide}.

\section{Orchestrating Multi-GPU PIR Execution}
\label{sec:multi-gpu}

\begin{figure}
    \centering
    \includegraphics[width=0.99\columnwidth]{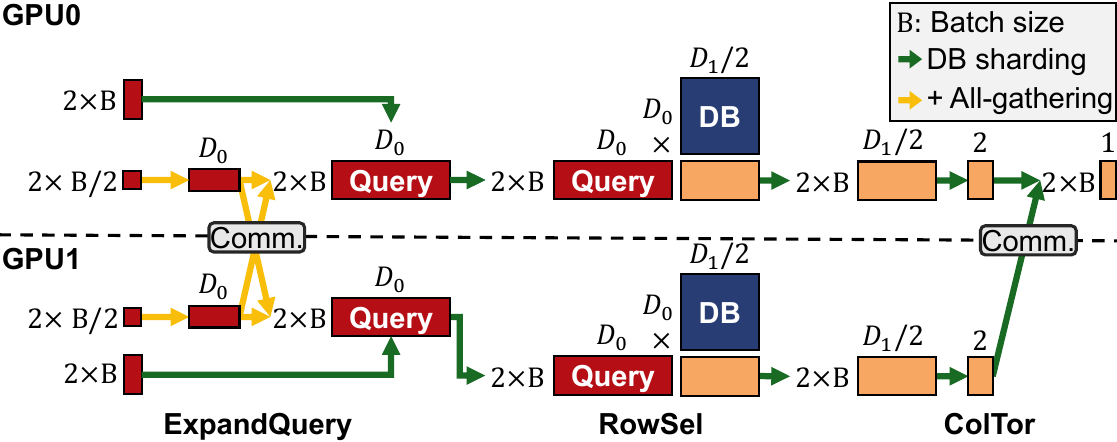}
    \vspace{-0.03in}
    \caption{\DB sharding and all-gathering expanded \ciphertexts in multi-GPU PIR execution on two GPUs. Comm. indicates inter-GPU communication.}
    \label{fig:multi-gpu-process}\Description{}
    \vspace{-0.03in}
\end{figure}

We investigate multi-GPU execution to scale PIR systems along two dimensions: increasing query throughput (queries per second, QPS) and supporting larger \DB capacity.
We consider three strategies: na\"ive batch parallelism, \DB sharding with response aggregation, and all-gathering expanded \ciphertexts.
These approaches serve different roles depending on whether the \DB fits within the DRAM capacity of a single GPU and whether high-bandwidth inter-GPU communication is available.
Fig.~\ref{fig:multi-gpu-process} shows \NAME{}’s multi-GPU execution flow and communication points; we denote the number of GPUs as $n_\text{GPU}$.

\subsubsection*{Na\"ive batch parallelism}
Na\"ive batch parallelism aims solely at increasing query throughput by evenly distributing a batch of queries across GPUs, with each GPU processing its assigned queries on a full copy of the \DB. 
This achieves near-linear scaling in QPS with $n_\text{GPU}$ without inter-GPU communication, but only while the \DB fits within a single GPU’s DRAM.

In practice, the supported \DB size is much smaller than DRAM capacity.
In OnionPIR, each \DB record is stored as a polynomial of $4N$ 32-bit elements (64\,KB for $N=2^{12}$) with NTT already applied to facilitate RowSel computations.
However, the information density of a plaintext polynomial is only $N\log P$ bits for a plaintext modulus $P$ (e.g., 16\,KB for $P=2^{32}$).
This results in a $4\times$ storage overhead, limiting the 32\,GB DRAM of an RTX~5090 to \DB sizes below 8\,GB.

\subsubsection*{\DB sharding with response aggregation}
\DB sharding partitions \DB along the $D_1$ dimension across $n_\text{GPU}$ GPUs (Fig.~\ref{fig:multi-gpu-process}), so each device stores $D_0\times \frac{D_1}{n_\text{GPU}}$ records, allowing a $n_\text{GPU}\times$ larger total \DB capacity.
Each GPU performs RowSel and ColTor on its local \DB shard and produces partial results that are merged into the final response.
Since each GPU sends only a single \ciphertext per query after ColTor (the right side of Fig.~\ref{fig:multi-gpu-process}),
the communication overhead remains modest even on PCIe-based systems.
ExpandQuery is executed on every GPU to provide the expanded \ciphertexts consumed by RowSel and ColTor.

Unlike prior work~\cite{hpca-2026-ive}, which uses \DB sharding to overcome capacity limitations, \NAME leverages it even for smaller \DB to boost throughput.
By distributing memory accesses across multiple GPUs, \NAME alleviates bottlenecks in bandwidth-critical phases.

\subsubsection*{All-gathering expanded \ciphertexts}
Building on \DB sharding, we exploit an all-gather for ExpandQuery to further enhance throughput.
Instead of running ExpandQuery redundantly on every GPU, we exploit na\"ive batch parallelism for ExpandQuery to all-gather expanded \ciphertexts.
After each GPU expands $\text{batch}/n_\text{GPU}$ queries, the results are fully shared across the GPUs such that each GPU holds the entire set of expanded \ciphertexts for the batch (see the left side of Fig.~\ref{fig:multi-gpu-process}).

Unlike response aggregation, this all-gather requires sharing $\text{batch}\cdot D_0$ \ciphertexts across the GPUs.
This high communication overhead can be mitigated by high-bandwidth interconnects.
For a batch size of 32, the data transfer can reach 1\,GB ($32 \cdot D_0 \cdot 128$\,KB for $D_0=256$).
Modern high-bandwidth interconnects such as NVLink provide hundreds of GB/s (e.g., 600--900\,GB/s~\cite{nvidia-nvlink,tpds-2020-nvlink}), rendering this overhead negligible compared to computation and DRAM access costs. 
Consequently, this approach linearly scales ExpandQuery throughput with minimal communication overhead.
\section{Evaluation}
\label{sec:eval}

\subsection{Experimental Setup}
We used cryptographic parameters providing 128-bit security~\cite{iacr-2024-guideline}.
Unless otherwise specified, experiments were conducted on an RTX~5090 with batch size 32.
We evaluated \DB sizes of 1\,GB, 2\,GB, and 4\,GB with 16\,KB records to capture a wide range of working-set and memory-traffic regimes.
All implementations were written in CUDA C++ and compiled with identical settings (nvcc 13.1.115) under the same GPU driver (NVIDIA Driver 590.48.01) for a fair comparison.
As our system assumes multi-client batching, we assess performance using query throughput (queries per second, QPS) and its reciprocal, amortized execution time per query, as the primary metrics.

For comparison, we developed a baseline GPU implementation with multi-client batching based on PIRonGPU~\cite{github-PIRonGPU, iacr-2024-HEonGPU} and IVE~\cite{hpca-2026-ive}.
Unless noted otherwise, reported results include only server-side execution time and memory traffic.
For multi-GPU experiments, we evaluated two-GPU systems based on RTX~5090 and four-GPU systems based on H100.
The RTX 5090 system uses PCIe 5.0 (64 GB/s), while the H100 system supports NVLink 4.0 (900 GB/s)~\cite{nvidia-nvlink}.

\subsection{Execution Time Enhancements}

\begin{figure}
    \centering
    \includegraphics[width=0.99\columnwidth]{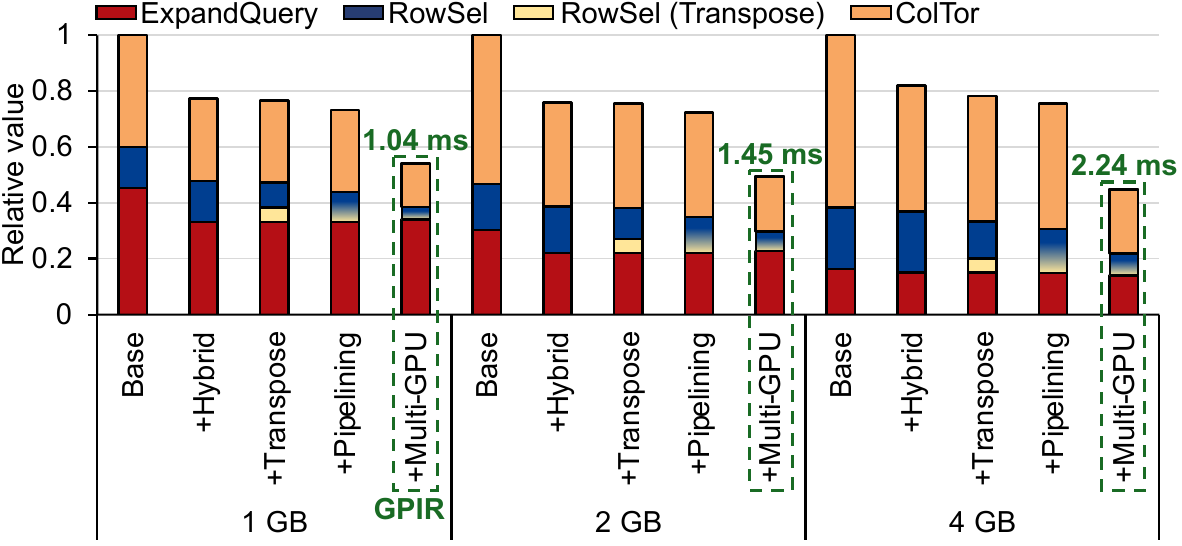}
    \vspace{-0.03in}
    \caption{Ablation study of our optimizations across different \DB sizes on an RTX 5090. We measured amortized execution time per query, while incrementally applying stage-aware hybrid execution (+Hybrid), transposed-layout RowSel (+Transpose), $p$-dimension pipelining (+Pipelining), and multi-GPU execution (+Multi-GPU).}
    \label{fig:ablation-study}\Description{}
    \vspace{-0.03in}
\end{figure}

Fig.~\ref{fig:ablation-study} shows the end-to-end execution time breakdown as successive optimizations are applied to the PIR computation.
Overall, \NAME achieves 1.84--2.23$\times$ speedups over the baseline (\textit{Base}) by combining stage-aware hybrid kernel execution (\textit{+Hybrid}), transposed-layout RowSel (\textit{+Transpose}), $p$-dimension pipelining (\textit{+Pipelining}), and multi-GPU execution (\textit{+Multi-GPU}).

Stage-aware hybrid kernel execution reduces latency by eliminating intermediate global memory traffic across kernel boundaries, yielding up to 1.37$\times$ speedups in ExpandQuery and 1.42$\times$ speedups in ColTor.
Applying transposed-layout RowSel significantly reduces the raw GEMM execution time by 1.50--1.65$\times$ through GEMM-friendly memory access patterns.
However, improvements in the total RowSel time, which includes both the transpositions and the GEMMs, are more modest at 1.03--1.20$\times$, as the additional transpose kernels account for up to 6.73\% of end-to-end runtime.
To address this, $p$-dimension pipelining overlaps transpositions with GEMMs, improving RowSel speedup to 1.28--1.41$\times$ over the baseline and establishing \NAME as the fastest configuration across all \DB sizes.

Finally, multi-GPU execution with two GPUs yields a further 1.35$\times$--1.68$\times$ end-to-end speedup over the pipelined configuration.
\NAME achieves an amortized execution time per query of 1.04\,ms, 1.45\,ms, and 2.24\,ms for 1\,GB, 2\,GB, and 4\,GB {\DB}s, respectively.
A detailed multi-GPU execution analysis is presented in \S\ref{sec:eval:scalability}.

\subsection{DRAM Traffic Reductions}

\begin{figure}
    \centering
    \includegraphics[width=0.99\columnwidth]{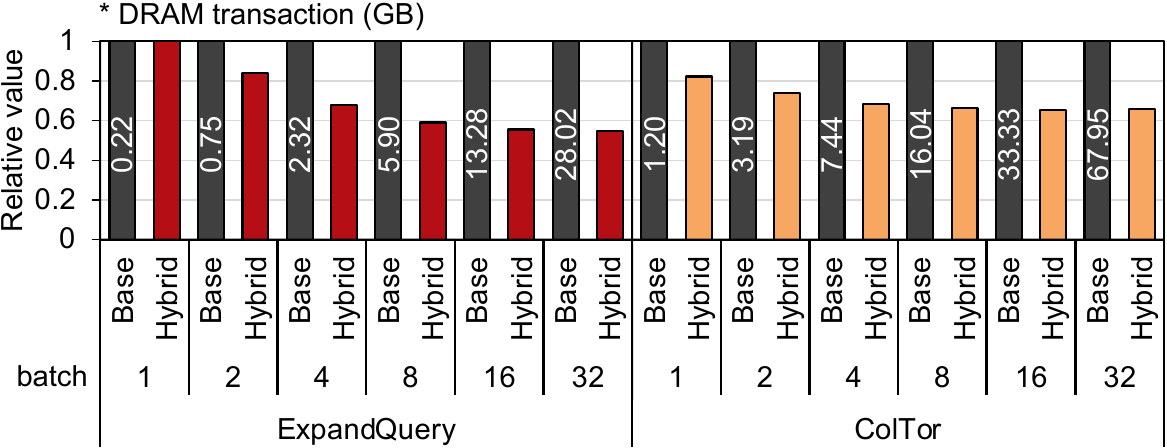}
    \vspace{-0.03in}
    \caption{DRAM transactions under different batch sizes for ExpandQuery and ColTor. Results are normalized to the baseline (operation-level kernels).}
    \label{fig:dram-access-per-batch}\Description{}
    \vspace{-0.03in}
\end{figure}

The stage-aware hybrid execution strategy consistently reduces DRAM transactions for both ExpandQuery and ColTor, and the benefit grows with batch size.
As shown in Fig.~\ref{fig:dram-access-per-batch}, at a batch size of 32, hybrid execution reduces DRAM transactions by up to 1.83$\times$ for ExpandQuery and 1.52$\times$ for ColTor compared to the baseline implementation (operation-level kernels).

This reduction is primarily enabled by applying stage-level kernels only at stages where the working set exceeds L2 capacity.
As a result, short-lived intermediate data are consumed on-chip instead of being spilled to DRAM, leading to lower DRAM traffic compared to the baseline execution.
The benefit becomes more pronounced as batch size increases, where transient working-set growth is amplified.
This reduction in DRAM traffic directly aligns with the execution-time speedups observed in Fig.~\ref{fig:ablation-study}.

\subsection{Evaluation across GPU Architectures}
\begin{figure}
    \centering
    \includegraphics[width=0.99\columnwidth]{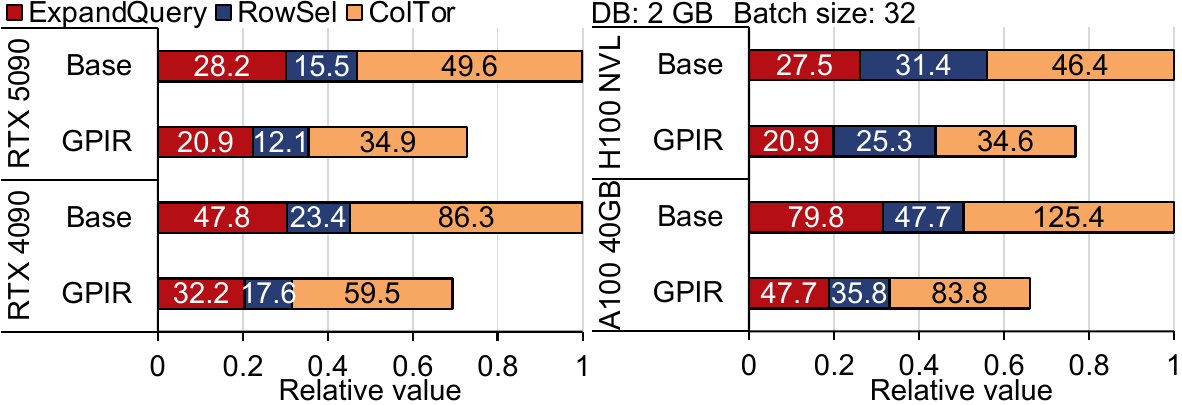}
    \vspace{-0.03in}
    \caption{Latency breakdown across GPUs for baseline and \NAME on a single GPU. Results are normalized to the baseline.}
    \label{fig:additional_gpu}\Description{}
    \vspace{-0.03in}
\end{figure}

Fig.~\ref{fig:additional_gpu} evaluates \NAME across multiple GPU architectures to demonstrate its generalizability. We extend our evaluation beyond RTX~5090 (L2 cache size: 96\,MB, DRAM bandwidth: 1.79\,TB/s) to RTX 4090 (72\,MB, 1.01\,TB/s), H100 NVL (50\,MB, 3.94\,TB/s), and A100 40GB (40\,MB, 1.56\,TB/s), covering a wide range of memory bandwidths and L2 cache capacities.

In ExpandQuery and ColTor, \NAME consistently improves performance by reducing DRAM traffic, achieving 1.33$\times$/1.42$\times$ speedups on RTX 5090 and comparable gains on H100 (1.32$\times$/1.34$\times$). Larger improvements are observed on GPUs with smaller L2 caches, such as RTX 4090 (1.48$\times$/1.45$\times$) and A100 (1.67$\times$/1.46$\times$), where DRAM traffic reduction becomes more critical.
This trend is driven by \NAME's stage-aware hybrid execution, which selects kernel granularity based on the relationship between working set size and L2 capacity. As L2 capacity decreases, the switching point shifts earlier (e.g., fourth stage on RTX~5090, fifth on RTX~4090, and sixth on A100), consistently matching execution to the dominant bottleneck (\S\ref{sec:opt-1:hybrid_kernel}).

In RowSel, \NAME achieves consistent speedups of 1.24--1.33$\times$ across all GPUs, confirming the NTT--GEMM data layout conflict as a fundamental constraint effectively mitigated across architectures.

\subsection{Multi-GPU Scalability}
\label{sec:eval:scalability}
\begin{figure}
    \centering
    \includegraphics[width=0.99\columnwidth]{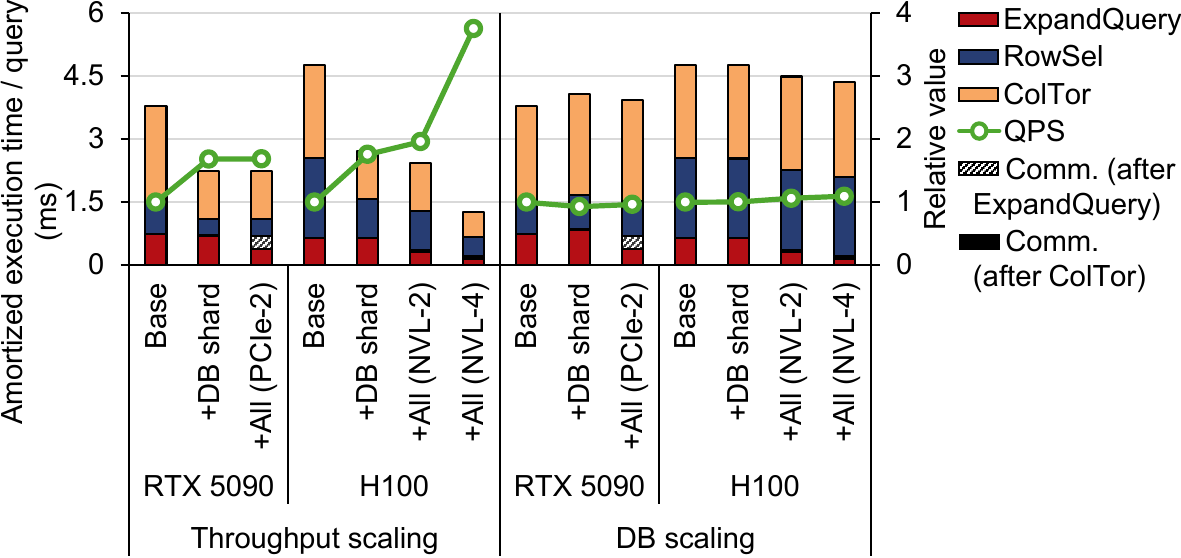}
    \vspace{-0.01in}
    \caption{Performance under throughput scaling (4\,GB \DB in total) and \DB scaling (4\,GB \DB per GPU). PCIe-2 uses two GPUs over PCIe, while NVL-2 and NVL-4 use two and four NVLink-connected GPUs, respectively. We evaluate \DB sharding (\textit{+\DB shard}) and additional all-gather of expanded \ciphertexts (\textit{+All}).}
    \vspace{-0.01in}
    \label{fig:multi-gpu}\Description{}
\end{figure}

Fig.~\ref{fig:multi-gpu} evaluates \NAME's multi-GPU scalability either to enhance PIR throughput at a fixed \DB size (throughput scaling) or to support $n_\text{GPU}=2,4$ times larger \DB sizes (\DB scaling).
The baseline configurations (\textit{Base}) use a 4GB \DB stored on a single GPU and represent our fully optimized single GPU implementation.
We incrementally applied \DB sharding (\textit{+DB shard}) and the all-gather of expanded \ciphertexts over PCIe (\textit{+All (PCIe)}) or NVLink (\textit{+All (NVL)}) if applicable.
We explicitly separate the communication overheads into two parts: after ExpandQuery and after ColTor, which differ substantially in both data volume and performance impact.

\noindent
\textbf{Throughput scaling:}
When scaling throughput by \DB sharding and all-gather in ExpandQuery, \NAME achieves substantial QPS improvements.
Our optimizations together result in a 1.69$\times$ speedup on RTX 5090 (PCIe).
For H100 (NVLink), \NAME achieves near-linear scaling, reaching 1.96$\times$ speedup with two GPUs and 3.76$\times$ speedup with four GPUs connected by a high-bandwidth interconnect.
The all-gather after ExpandQuery incurs noticeable overheads for PCIe-based communication, accounting for up to 7.96--13.9\% of the total execution time, due to its relatively large data volume (Fig.~\ref{fig:multi-gpu-process}).
However, this data volume does not increase with larger \DB sizes or more GPUs, as it scales only with the fixed parameters of batch size and $D_0$ (\S\ref{sec:multi-gpu}).
On systems with NVLink, the same communication is almost entirely hidden, contributing only 0.70--2.46\% of the runtime.
In contrast, the communication after ColTor aggregates tiny partial tournament results and remains negligible across all configurations, accounting for at most 0.26\% even on PCIe.

\noindent
\textbf{DB scaling:}
When the total \DB capacity increases from 4\,GB to 8\,GB by sharding the \DB across two GPUs, \NAME maintains comparable throughput, achieving 0.96$\times$ and 1.06$\times$ QPS on RTX 5090 and H100, respectively. On H100, further scaling the \DB to 16\,GB across four GPUs even yields a 1.09$\times$ QPS improvement.
These results demonstrate that \NAME can scale \DB capacity without incurring proportional performance degradation, enabled by efficient \DB sharding and low-overhead inter-GPU communication.
Notably, since ExpandQuery generates a fixed number of $D_0$ expanded \ciphertexts per query regardless of \DB size, increasing the \DB capacity introduces only marginal extra work to this stage, which can even lead to a slight throughput increase (up to 1.09$\times$) due to improved inter-GPU parallelism.

\subsection{Comparison with prior work}

\setlength{\tabcolsep}{4pt}
\begin{table}[t]
    \centering
    \caption{Throughput (QPS) and speedup across \DB sizes for \NAME and prior GPU-based PIR proposals on an RTX~5090.}
    \label{tab:qps_comparison}
    \vspace{-0.05in}
{
    \small
    \begin{tabular}{l|cc|cc|ccc}
    \toprule
    \DB & PIRon & Shift & \NAME & \NAME  & \NAME  & vs.\ PIRon & vs.\ Shift \\
    size & GPU & PIR* & (S)$\dagger$ & (B)$\dagger$ & (M)$\dagger$ & GPU & PIR  \\
    \midrule
    1\,GB & 5.3 & 12.5 & 185.8 & 709.8 & 958.7 & 180.9$\times$ & \phantom{0}76.7$\times$ \\
    2\,GB & 2.8 & - & 118.4 & 473.6 & 690.3 & 246.5$\times$ & - \\
    4\,GB & 1.5 & \phantom{0}3.6 & \phantom{0}69.4 & 264.2 & 445.8 & 297.2$\times$ & 123.8$\times$ \\
    \bottomrule
    \end{tabular}
}
    \vspace{4pt}
    {\footnotesize
    \begin{itemize}[leftmargin=*, nosep]
    \item[*] Reported values from the original paper (measured on an RTX 4090) due to unavailable open-source implementation.
    \item[$\dagger$] S denotes the single-batch (single-client) configuration, B denotes the multi-client batching configuration, while M denotes the multi-GPU execution with two GPUs.
    \end{itemize}
    }
\end{table}
\setlength{\tabcolsep}{6pt}

Table~\ref{tab:qps_comparison} shows that \NAME consistently delivers orders-of-magnitude higher throughput than prior GPU-based PIR systems across all evaluated \DB sizes.
We compare only single-server, GPU-based PIR systems without restrictive assumptions discussed in \S\ref{sec:background:pir}.

PIRonGPU~\cite{github-PIRonGPU} and ShiftPIR~\cite{ccs-2025-shiftpir} support only single-client (non-batched), single GPU execution.
Even under this setting, \NAME (S) already achieves substantial performance gains through a carefully optimized GPU implementation, outperforming PIRonGPU by up to 46.3$\times$ and ShiftPIR by up to 19.3$\times$.
Multi-client batching further amplifies these gains, with \NAME (B) achieving up to 176.8$\times$ speedups over PIRonGPU and 73.7$\times$ over ShiftPIR, while sustaining high throughput as the \DB size increases.
Finally, the multi-GPU configuration, \NAME (M), extends these benefits by scaling query throughput across two GPUs, achieving up to a 297.2$\times$ improvement over the prior state-of-the-art open-source implementation.

These results show that \NAME maintains high throughput as the \DB size grows not simply by batching queries, but by redesigning GPU execution to match the dominant bottleneck of each PIR phase.

\section{Related Work}
\label{sec:related}
\subsubsection*{Other PIR protocols}
A large body of prior work enhances the practicality of PIR by reducing either server computation or communication costs.
Early efforts focus on communication efficiency by exploiting algebraic properties of HE.
FastPIR~\cite{osdi-2021-fastpir} and its refinement, INSPIRE~\cite{isca-2022-inspire}, reduce query expansion and server computation via automorphisms.
Complementary approaches~\cite{ccs-2021-onionpir, sp-2022-spiral, ccs-2024-respire, security-2021-mulpir,sp-2018-sealpir} further reduce communication through query compression and external products, often at the cost of increased server-side computation.
Other studies~\cite{usenixsec-2023-simplepir, sp-2024-piano} reduce online computation by introducing offline preprocessing at the cost of additional client-side storage or setup.
More recent designs~\cite{iacr-2025-inspire, sp-2026-via} aim to jointly reduce communication and server computation by balancing these trade-offs.

\subsubsection*{Hardware acceleration of PIR}
Several studies explore hardware acceleration to meet increasing computational demands. INSPIRE~\cite{isca-2022-inspire} proposes an SSD-based in-storage accelerator, while SmartPIR~\cite{micro-2025-smartpir} and Conflux~\cite{hpca-2026-conflux} extend this to computational storage devices (CSDs), reducing data movement through specialized hardware.
IVE~\cite{hpca-2026-ive} proposes a custom accelerator for OnionPIRv2~\cite{iacr-2025-onionpirv2}, showing that multi-client batching improves throughput but introduces data-movement bottlenecks. 
It addresses these challenges using large on-chip scratchpads and tightly scheduled execution, highlighting the importance of hardware support for data movement and reuse.

\subsubsection*{GPU acceleration of PIR}
Several studies accelerated PIR using GPUs under different threat models and protocol assumptions.
Lam et al.~\cite{ASPLOS-2024-gpupir} targets two-server PIR built on distributed point functions (DPFs), inheriting the multi-party computation model and exploiting GPU parallelism to accelerate DPF evaluations.
CIP-PIR~\cite{usenixsec-2022-gpupir} similarly focuses on multi-server or information-theoretic PIR schemes, leveraging massive parallelism to improve throughput.
ShiftPIR~\cite{ccs-2025-shiftpir} offloads query processing to GPUs and addresses GPU memory limitations by utilizing host memory, but is limited by host--device bandwidth due to full-database scans.
Zero-copy and unified virtual memory can further extend effective memory capacity, but do not eliminate this limitation. 
Nevertheless, leveraging host memory may still serve as a practical fallback when the database exceeds GPU capacity. 
In particular, tightly integrated CPU--GPU systems with higher host--device bandwidth (e.g., NVIDIA Grace Hopper Superchip~\cite{nvidia-gh200}) could help mitigate this limitation.
\section{Conclusion}
\label{sec:conclusion}

This paper presents \NAME, a GPU-accelerated PIR system that addresses architectural challenges in making lattice-based single-server PIR practical under multi-client batching.
While multi-client batching is necessary for high throughput, it introduces substantial working-set expansion in ExpandQuery and ColTor and layout conflicts in RowSel that significantly constrain performance.
By addressing these challenges with stage-aware hybrid execution, transposed-layout RowSel, $p$-dimension pipelining, and multi-GPU execution, \NAME achieves up to a 2.23$\times$ speedup over the baseline implementation and up to 297.2$\times$ higher throughput than the state-of-the-art open-source GPU-based PIR system.
Further, \NAME scales efficiently in multi-GPU environments, improving QPS by up to 1.69$\times$ on RTX 5090 and 1.96$\times$ on NVLink-connected H100 with two GPUs under throughput scaling, and reaching 3.76$\times$ on H100 with four GPUs.
Under \DB scaling, it maintains near-constant QPS (0.96$\times$ on RTX 5090 and 1.09$\times$ on H100) even when the \DB no longer fits in a single GPU’s memory.
Overall, architecture-aware software designs can make lattice-based PIR a practical solution for large-scale, privacy-preserving database services on modern accelerator platforms.

\begin{acks}
\begin{sloppypar}
We would like to express our gratitude to Sangpyo Kim, Yue Chen, and Ling Ren for the numerous productive discussions concerning the implementation of OnionPIRv2.
This work was supported by an Institute of Information \& communications Technology Planning \& Evaluation (IITP) grant funded by the Korea government (MSIT) (RS-2021-II211343, RS-2025-02217656, and RS-2025-02304125).
Hyesung Ji and Jongmin Kim are with the Interdisciplinary Program in Artificial Intelligence (IPAI), SNU.
Jung Ho Ahn, the corresponding author, is with the Department of Intelligence and Information and IPAI, SNU.
\end{sloppypar}
\end{acks}

\bibliographystyle{ACM-Reference-Format}
\bibliography{ref}

\end{document}